Discussion Paper:Communication based on unilateral preference on Twitter: Internet luring in Japan(2018)

# Deciphering Unilateral Communication Patterns in Directed Temporal Networks: Network Role Distribution Approach

Yasuko Kawahata [†]

Faculty of Sociology, Department of Media Sociology, Rikkyo University, 3-34-1 Nishi-Ikebukuro,Toshima-ku, Tokyo, 171-8501, JAPAN.
`ykawahata@rikkyo.ac.jp,kawahata.lab3@damp.tottori-u.ac.jp`

**Abstract:** In the vast expanse of online communication, identifying unilateral preference patterns can be pivotal in understanding and mitigating risks such as predatory behavior. This paper presents a comprehensive approach to dissect and visualize such patterns in social networks. Through the lens of a directed network model, we simulate a scenario where a predominant cluster $A$ disperses information unilaterally towards a much larger, but passive, cluster $B$, while being overseen by a vigilant cluster $C$, restricted by an information blocking cluster $D$, and countered by an alerting cluster $E$. The key novelty of our approach lies in the integration of dynamic $k$-core analysis to reveal the structural robustness and core areas of influence within the network. By simulating the network over 1000 time steps, we trace the evolution of message flows and the emergence of core-periphery structures. The $k$-core visualization offers a vivid depiction of the network's resilience and the centrality of different clusters, providing a macroscopic view of the network's topology. Our methodology captures the expansion of the largest connected component and identifies pivotal nodes that may act as gatekeepers or influencers within the network. The findings emphasize the potential of $k$-core analysis in augmenting the detection of unidirectional communication patterns, offering insights into the underlying architecture that facilitates such interactions. This study contributes to the domain of network analysis by showcasing a novel application of $k$-core decomposition in the context of social network simulations, which can be instrumental in safeguarding online communications.

## 1. Introduction

In the vast expanse of online communication, identifying unilateral preference patterns can be pivotal in understanding and mitigating risks such as predatory behavior. This paper presents a comprehensive approach to dissect and visualize such patterns in social networks. Through the lens of a directed network model, we simulate a scenario where a predominant cluster $A$ disperses information unilaterally towards a much larger, but passive, cluster $B$, while being overseen by a vigilant cluster $C$, restricted by an information blocking cluster $D$, and countered by an alerting cluster $E$. Incorporated into this study is a simulation framework that models the flow of information across a directed network comprising various clusters with distinct roles and communication behaviors. The simulation employs a dynamic system where clusters $A$ through $E$ interact over a series of time steps, with each cluster's activity shaped by both intrinsic message-generation rules and external media influences.

The network's initialization reflects a realistic distribution of roles among the clusters: Cluster $A$ as a prolific information sender; Cluster $B$ as the primary recipient; Cluster $C$ as the overseer and disseminator; Cluster $D$ as the gatekeeper, and Cluster $E$ as the reactive alarm-raiser. This setup is encoded in the `initialize_network` function, which sets up the directed graph and assigns each node to its respective cluster.

A key feature of our model is the introduction of media influence, which exerts variable impacts on the clusters. This is represented by two distinct dictionaries, `amedia` and `bmedia`, that store the positive and negative media influence values, respectively. The influence values are not static; they are subject to random fluctuations, as encapsulated by the `randomize_media_values` function, which imbues the model with a degree of unpredictability and realism.

The `update_network` function is pivotal to the simulation, dictating the propagation of messages from nodes in Cluster $A$ to other clusters, and the potential cessation of activity when certain thresholds are met. Cluster $E$ plays a critical role here, as it can issue alerts to Cluster $A$, with a probability of silencing $A$ nodes that become overly inundated with messages. This mechanism serves to mimic the regulatory feedback that occurs in real-world networks, where excessive message dissemination can trigger countermeasures.

The simulation is visualized through functions like `plot_degree_distribution`, which illustrates the distribution of connections at various time steps, offering insight into the evolving structure of the network. The degree distribution plots and time series of the clusters' influence levels give a quantitative backbone to the qualitative analysis, allowing us to observe the emergent properties and trends within the network over time.

By tracking the accumulated media influence on each cluster, we gain a nuanced understanding of the long-term effects of media on communication patterns. The results provide a window into the cyclical nature of influence and the propagation of information, with potential applications in detecting and mitigating unilateral communication patterns that could signal harmful activities such as online predation.

This study, therefore, presents a comprehensive approach that combines network theory, simulation modeling, and dynamic media influence analysis to explore and understand the complexities of unilateral preference communication within social networks.

## 2. Previous Research

### 2.1 Reactive Alarm Raiser Adversarial Networks

Adversarial Networks is a concept found primarily in cybersecurity, fraud detection, strategic game theory, and other fields. Adversarial network approaches are used to analyze the interaction between entities with conflicting objectives; Justin, S., et al. (2020) discuss an approach to detect cybersecurity breaches using deep learning. Adversarial networks were often used to model the interaction between attackers and defenders; Zenati, H., et al. (2018) applied GANs to anomaly detection tasks through adversarial processes between normal and anomalous data; Roy, S., et al. (2010) et al. suggested that game theory can be used to analyze the dynamics between the different strategies of attackers and defenders; Biggio, B., Roli, F. (2018) et al. suggested that adversarial machine learning can be used to understand how models behave in response to deliberate attacks They gave an example of a case study they studied in order to Studies combine game theory and Adversarial Networks to model interactions between adversarial agents and find optimal strategies. These studies explore ways to analyze game-theoretic interactions between agents with adversarial behavior in a variety of network environments and situations. Each study combines theoretical insights with practical applications to shed light on network security and strategic decision making issues. The work of Alpcan, T., Başar, T. (2003) proposes a game-theoretic approach for network breach detection and analyzes the They analyze strategic interactions; Nguyen, K. C., Alpcan, T., Başar, T. (2009) in their work model and analyze security problems in networks with interdependent nodes using multistage stochastic games; Tambe, M. (2011) provides a broad perspective on the theory of security games and its application to real systems, analyzing adversarial interactions between agents; Korilis, Y. A., Lazar, A. A., and Orda, A. (1997) use Stackelberg game theory to Hausken, K. (2013) discussed the use of game theory in the context of cyber warfare and proposed modeling attack and defense strategies. In addition, research on fraud detection typically focuses on developing methods to identify and defend against hostile behavior. Below are some examples of studies that incorporate fraud detection and adversarial network approaches. A variety of techniques are utilized to improve fraud detection and include deep learning, anomaly detection algorithms, game theory, and adversarial network principles. These are intended to improve the accuracy of fraud detection and increase the ability to respond in real time. Dal Pozzolo, A., Boracchi, G., Caelen, O., Alippi, C., Bontempi, G. (2018) proposed a deep learning-based model for detecting credit card fraud and proposed content for real-time fraud detection Buczak, A. L., Guven, E. (2016) investigated data mining and machine learning methods in cybersecurity intrusion detection and explored approaches to fraud detection in hostile environments. Alpcan, T., Başar, T. (2006) uses game theory to model intrusion detection systems in environments with limited observations and analyze optimal defense strategies. Schlegl, T., Seebock, P., Waldstein, S. M., Schmidt-Erfurth, U., Langs, G. (2017) uses Generative Adversarial Networks (GANs) to detect unmonitored anomalies, particularly assist in the discovery of markers in medical images. Mirsky, Y., Doitshman, T., Elovici, Y., Shabtai, A. (2018) proposed a method for intrusion detection in online networks using an ensemble of autoencoders and proposed identifying anomalous traffic patterns.

## 3. Network Building

In this method, we aim to generate and simulate a dataset characterized by the attributes $A, B, C, D$, and $E$. Each cluster has distinct behaviors in terms of information exchange:

> $A$ is a cluster that unilaterally sends information.
>
> $B$ is a cluster that unilaterally receives information.
>
> $C$ is a cluster that observes the exchange between $A$ and $B$ and disseminates the information.
>
> $D$ is a cluster that blocks information from $A$, $B$, and $C$.
>
> $E$ is a cluster that issues alerts to $A$ in an attempt to halt its unilateral information dissemination.

The population size order is $B > D > A > C > E$. Additionally, the volume of information exchange is presumed to be $A > E$, $E > A$, and $C > B > D$.

Cluster $A$ unilaterally sends information, implying that a member of $A$ has directed edges to members of other clusters. Cluster $B$ only receives information, indicated by having incoming edges without outgoing ones. Cluster $C$ both receives from $A$ and propagates information, while $D$ has no incoming edges, and $E$ specifically targets $A$ with alerts.

The objective is to identify the largest connected components of the network and visualize $k$-cores, where each node in a $k$-core has at least $k$ connections. We consider communication patterns beyond unilateral exchanges.

The behavior of the clusters over timesteps is subject to the following rules:

- $A$: Continuously sends messages unilaterally.
- $B$: Responds to some of $A$'s messages.
- $C$: Observes and disseminates information.
- $D$: Strengthens information blocking.
- $E$: Aims to disrupt $A$'s network by sending unilateral alerts.

If an agent in $A$ receives more than 15 messages in a timestep, it will cease its activities. The model construction is described as follows.

### 3.1 Building of Negative Media, Positive Media Effects

We need to update the model so that the $E$ cluster behaves in such a way that it forms a hostile network against $A$ with unilateral messaging and spreads its influence to the other clusters ($B, C, D$). Based on the requirements here, we update the model as follows: $E$ sends information unilaterally to $A$, $A$ ignores information from $E$, but if more than 15 pieces of information come to one $A$ node during a timestep, that $A$ node stops acting at that timestep. $E$ sends out negative information about $A$ to clusters $B, C$, and $D$. The effect is to decrease $A$'s ability to transmit information by 10 during the timestep (i.e., as negative information from $E$ spreads to $B, C$, and $D$, $A$'s ability to transmit decreases).

In order to incorporate the effects of the external influence variables positive media amedia($timesteps$) and negative media bmedia($timesteps$) into the network, the effect of each media on each cluster must be defined and reflected in the update function. amedia is set to decrease the positive effect for cluster $A$ with one-way messaging, and bmedia is set to increase the negative effect for cluster $A$ with one-way messaging. Each cluster plays a special role in amedia to reduce the positive messages for $A$ and bmedia to increase the negative impact for $A$.

Let sizes be a dictionary representing the size of each cluster in the network. Let messages be a dictionary representing the initial number of messages for each cluster. For a network $G$, each cluster $c \in \{A, B, C, D, E\}$ is initialized with $N_c$ = sizes[$c$] nodes and starts with $M_c(0)$ = messages[$c$] messages.

The apply_media_influence function updates the message count for each cluster based on the media influence and the current timestep $t$. For each cluster $c$, the updated number of messages $M'_c(t)$ after applying media influence at timestep $t$ is given by:

$$M'_c(t) = \max(M_c(t-1) + (S_c \cdot t), 0)$$

where $S_c$ is the initial media influence on cluster $c$, which is defined based on an initial condition:

$$S_c = \begin{cases} \text{amedia}[c] & \text{if initial condition is 'A'} \\ \text{bmedia}[c] & \text{if initial condition is 'B'} \end{cases}$$

For clusters not influenced by the selected media, $S_c$ is considered to be 0.

The influence of media on cluster $c$ scales linearly with time, such that the longer the media influence persists, the greater the cumulative effect on the message count for that cluster.

### 3.2 Building of a Cluster Network

The initial configuration of the clusters is as follows:

Cluster $A$: Size $n_A$, information sent $i_{\text{out},A}$, information received $i_{\text{in},A}$

Cluster $B$: Size $n_B$, information sent $i_{\text{out},B}$, information received $i_{\text{in},B}$

Cluster $C$: Size $n_C$, information sent $i_{\text{out},C}$, information received $i_{\text{in},C}$

Cluster $D$: Size $n_D$, information sent $i_{\text{out},D}$, information received $i_{\text{in},D}$

Cluster $E$: Size $n_E$, information sent $i_{\text{out},E}$ (a random variable), information received $i_{\text{in},E}$

The communication rules between clusters are as follows:

Cluster $A \rightarrow$ Cluster $B$ : Probability $p_{A \rightarrow B}$

Cluster $C \rightarrow$ Cluster $A$ : Probability $p_{C \rightarrow A}$,

Cluster $C \rightarrow$ Cluster $B$ : Probability $p_{C \rightarrow B}$

Cluster $E \rightarrow$ Cluster $A$ : Probability $p_{E \rightarrow A}$

Consider a directed graph $G = (V, E)$ representing the network, where $V$ is the set of nodes and $E$ is the set of edges.

Let sizes be a dictionary with the number of nodes in each cluster, and messages be a dictionary with the initial number of messages for each cluster. The timestep is denoted by $t$, and initial_condition determines the media influence.

At each timestep $t$, the media influence is applied to each cluster by updating the message counts according to the function apply_media_influence.

Let stop_acting be a boolean array indexed by nodes in cluster A, initially set to False. Cluster E sends alerts to cluster A, represented by a random number of messages $e_{\text{messages}}$ between 100 and 10000. For each node $e \in E$:

For each alert sent to a node $a \in A$, an edge is added from $e$ to $a$.

If the in-degree of $a$ reaches 15 or more, $a$ will stop acting, i.e., stop_acting$[a]$ = True.

The number of messages in cluster A is reduced by $e_{\text{messages}}$, ensuring it does not go below zero.

For each node $a \in A$ that is not stopped:

Messages are sent to other clusters $B, C, D$, and $E$. The number of messages sent is proportional to the number of messages in $A$ divided by the size of the target cluster, ensuring it's at least 1 to avoid division by zero.

For each message, an edge is added from $a$ to a randomly selected node in the target cluster.

Every 10 timesteps, the degree distribution of the graph $G$ is plotted to monitor the changes over time.

**Network Update Function**

Consider a directed graph $G = (V, E)$ where $V$ represents the set of vertices or nodes, and $E$ represents the set of directed edges between these nodes. Nodes are partitioned into clusters $A, B, C, D$, and $E$, with each cluster having a distinct role in the network's message passing dynamics.

Let sizes be a dictionary that maps each cluster to the number of its nodes, and messages be a dictionary that maps each cluster to its initial message count. The timestep is denoted by $t$, and initial_condition specifies the type of media influence to apply, with 'A' indicating positive influence and 'B' indicating negative influence. The variable $e\_influence$ is a dictionary that tracks the influence of cluster $E$ on the other clusters over time.

### 3.3 Media Influence Application

At each timestep $t$, the media influence dictated by initial_condition is applied:

$$\text{media\_influence} = \begin{cases} a_{\text{media}} & \text{if initial\_condition} =' A' \\ b_{\text{media}} & \text{if initial\_condition} =' B' \end{cases}$$

The message count for each cluster is then updated accordingly.

### 3.4 Interaction Between Clusters $E$ and $A$

For each node $e$ in cluster $E$, a random number of messages $e\_messages$ is sent to nodes in cluster $A$. If a node in $A$ receives 15 or more messages from $E$, it is marked to stop acting (i.e., stop sending messages).

### 3.5 Message Passing

Nodes in cluster $A$ that are not marked to stop acting will send messages to nodes in other clusters $B, C, D$, and $E$. The number of messages sent from a node in $A$ to each target cluster is proportional to the message count of $A$ divided by the size of the target cluster, ensuring at least one message is sent to avoid division by zero.

**Degree Distribution**

If $t$ is a multiple of 10, the degree distribution of graph $G$ is plotted to visualize the network's evolution.

### 3.6 Influence Tracking

The influence of cluster $E$ on each cluster is recorded by summing the in-degrees of all nodes in a given cluster. The simulation runs for 100 timesteps. At each timestep, the network is updated, and the influence of cluster $E$ is tracked.

**$E$ Cluster's Influence**

At the end of the simulation, the accumulated influence of cluster $E$ on each other cluster is visualized as a time series.

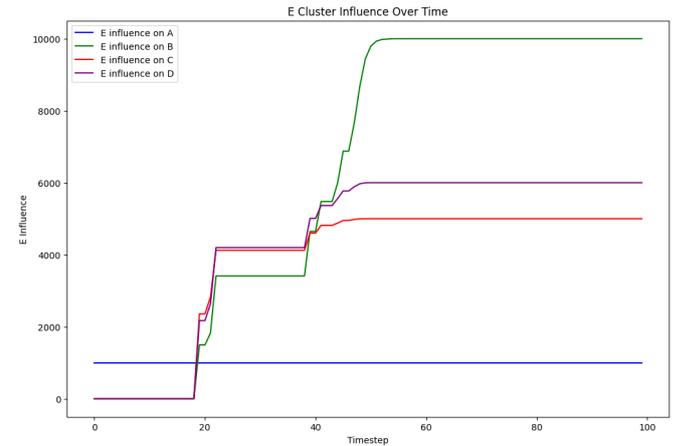

Fig. 1: E Cluster Influence Over Time

**Largest Connected Component**

Given a directed graph $G$, we define the following process for visualizing the largest connected component and its k-cores at a given timestep $t$ and for a given media type.

### 3.7 Connected Component and k-cores

Let $G_{cc}$ be the largest weakly connected component of $G$. Within $G_{cc}$, we identify k-cores for $k = 2, 6$, and 23, denoted as $K_2, K_6$, and $K_{23}$, respectively.

**Node Positioning**

The position of each node is determined using a force-directed layout algorithm (e.g., the spring layout). Each cluster $A, B, C, D,$ and $E$ is assigned a specific color for visualization, with clusters $A$ and $E$ having unique colors and the others colored based on the media type influencing them.

For each cluster $X \in \{A, B, C, D, E\}$, we plot the nodes of $X$ in the graph $G_{cc}$ using the assigned colors.

Nodes belonging to k-cores $K_2, K_6,$, and $K_{23}$ are drawn on top of the respective clusters with distinct colors to highlight the core structure.

Edges of the graph are plotted with a light transparency to emphasize the nodes and k-cores.

We calculate the centroid of each cluster in $G_{cc}$ to position the labels centrally.

A legend is created to differentiate between the clusters, k-cores, and media types.

Titled "Largest Connected Component with k-core (k=2, 6, 23) at Timestep $t$ type: media_type" and displayed without axes for a clean visualization.

## 4. Discussion

### 4.1 Update Network Function with timesteps

The `update_network` function modifies the directed graph $G$ based on the current state of messages within each cluster and the time step $t$.

### 4.2 Messages Handling

Let $\mathcal{M}$ be the message count for each cluster. We first ensure that each message count $\mathcal{M}_i$ is an integer value.

**Message Propagation**

For cluster $A$, if a node $a$ has fewer than 15 outgoing messages, it will send messages to nodes in clusters $B, C, D,$ and $E$. The number of messages sent by $a$ is determined by the integer division of $\mathcal{M}_A$ by the total number of potential target nodes.

**Alerts by Cluster $E$**

Cluster $E$ sends alerts to cluster $A$, which may stop nodes in $A$ from acting if they receive too many messages.

**Information Spread by Cluster $C$**

Cluster $C$ observes clusters $A$ and $B$ and spreads information to clusters $D$ and $E$.

**Information Blocking by Cluster $D$**

Cluster $D$ blocks information by not creating any outgoing edges.

Media values are randomized at each step.

Media influence is applied alternately, with positive influence at even and negative at odd time steps.

The network is updated based on the current message counts and the time step.

Every 10 time steps, the network is visualized, highlighting the largest connected component and the influence of media type.

**Mathematical Formulation**

Given a set of clusters $C = \{A, B, C, D, E\}$, sizes $\mathcal{S}$, and messages $\mathcal{M}$, the update rules at time step $t$ are as follows:

$$\mathcal{M}^{(t)} = \texttt{ensure\_integer}(\mathcal{M}^{(t-1)}+\texttt{media\_influence}(\mathcal{M}^{(t-1)},t))$$

$$\forall a \in A, \text{ if } \texttt{out\_degree}(a) < 15: \texttt{send\_messages}(a, \mathcal{M}_A, \mathcal{S})$$

$$\forall e \in E, \texttt{send\_alerts}(e, A)$$

$$\forall c \in C, \texttt{spread\_info}(c, \mathcal{M}_C, \mathcal{S})$$

$$\forall d \in D, \texttt{block\_info}(d)$$

Every time steps, the largest connected component is visualized with nodes colored according to the current media influence.

## 5. Conclusion

The resulting parameters discussed in the conclusion were: The following are the conditions for five classes A-E.

The conditions for the five classes A-E are summarized as follows:

**amedia effect**

| Class | Size | Messages | *amedia* |
|---|---|---|---|
| A | 100 | 200 | 100 |
| B | 100 | 100 | 50 |
| C | 50 | 100 | 10 |
| D | 60 | 50 | 10 |
| E | 10 | 100 | - |

Table. 1: Class conditions with media influences

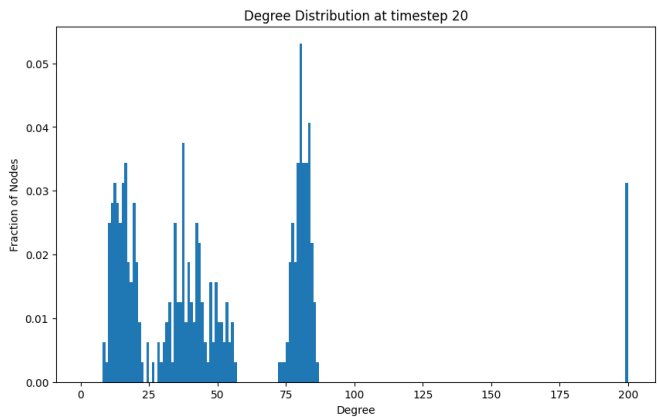

Fig. 2: Degree Distribution at timestep

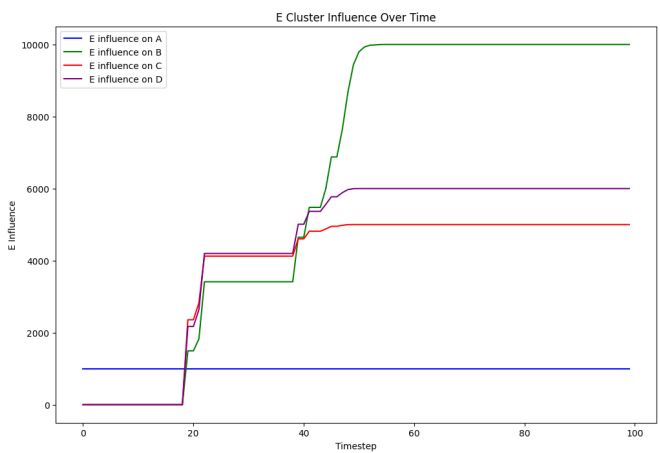

Fig. 3: E Cluster Influence Over Time

Note: Positive media influence (*amedia*) and negative media influence (*bmedia*) are shown to suppress information for each class. Classes without specified media influence are marked with a dash (-).

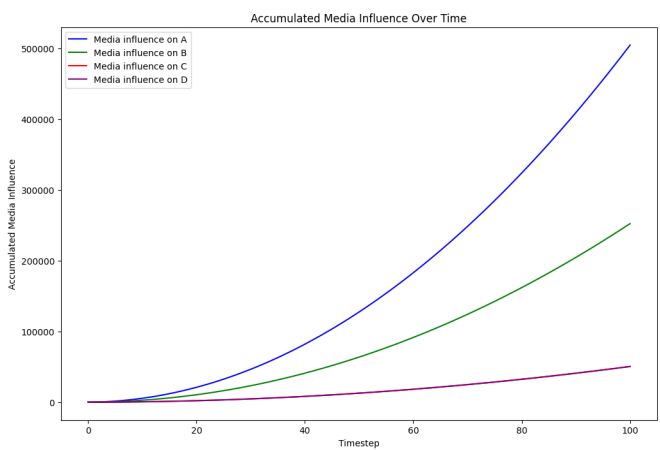

Fig. 4: Accumulated Media Influence Over Time.

The above results are for the case where only positive amedia are valid. The histogram (frequency distribution) shows the distribution of the degree of connectivity (number of edges) of the nodes in the network, with multiple peaks at time step $t = 20$. The histogram is a bit irregular and shows how the frequency distribution changes over time. At time step $t = 10$, most nodes have fairly low frequencies and a few have very high frequencies. This suggests that the network has a very heterogeneous connectivity at that point.

The $E$ cluster impact graph tracks the impact of the $E$ cluster on the other clusters ($A$, $B$, $C$, and $D$) over time. There is a spike in the impact of $E$ at certain time steps, which probably corresponds to events or changes within the network. For example, it could be the addition of a new edge or the acquisition of a critical node.

The last graph shows the cumulative impact of media on the cluster. It shows how the media's impact on clusters $A$, $B$, $C$, and $D$ increases over time, with the media's impact on clusters $A$ and $B$ being particularly significant.

These graphs provide many insights into the dynamics of the network, particularly the evolution of influence between clusters, the influence of the media, and the evolution of the distribution of network connectivity. These insights devise applications such as social network analysis, the spread of infectious diseases, information flows, or the dynamics of social influence.

## Network Dynamics and Cluster Behaviors

**(1) Considerations for clusters engaging in risky behavior, unilateral inductive behavior, and defensive behavior**

**Clusters engaging in risky behavior:** Cluster $A$ continues to be affected by $E$ at the lowest level and may be risk-averse or the most resistant to $E$'s influence.

**Clusters engaged in unilaterally guided behavior:** Cluster $B$ is influenced by $E$ at the highest level over time and may be more likely to be unilaterally guided by $E$'s intentions or opinions.

**Clusters acting defensively:** Clusters $C$ and $D$ began to be influenced along the way, but have stagnated at a certain level. This indicates a defensive attitude and may indicate a certain degree of resistance or adaptation to the influence of $E$.

**(2) Information selection behavior to be taken by passive cluster $B$**

Cluster $B$ can offset the influence of the $E$ cluster by increasing the diversity of its choices. For example, by paying attention to other information sources and viewpoints, a balanced opinion formation can be achieved. It is also important to reinforce a conscious decision-making process and evaluate how external influences affect their choices.

### (3) Consideration of social clusters and patterns of opinion formation:

The graph shows that cluster $E$ has a strong ability to influence other clusters, which may correspond to socially influential groups or opinion leaders. In opinion formation, it indicates that Cluster $B$ is more susceptible to influence and its opinions are more likely to be driven. This may indicate a group that is sensitive to trends and general social pressures.

### (4) Tendency of each cluster in terms of time trends

Cluster $A$ consistently has the lowest influence of $E$, which can be interpreted as having inherent values and beliefs and maintaining a strong stance against external influences. Cluster $B$ has increasing influence over time and appears to be very receptive to new information and trends. Clusters $C$ and $D$ show a sharp increase in influence at the beginning of their influence, but then remain stable at a certain level of influence. This may indicate that they adopt some degree of new information but then take a conservative stance.

### (5) Cluster $A$ - Independents:

Cluster $A$ is largely uninfluenced by $E$ and can, for example, be thought of as a group that holds to traditional values and adheres to its own beliefs without being influenced by outside trends or opinions. Socially, this might correspond to an independent professional, academic community, or a political group that strictly adheres to a particular philosophy or ideology.

### (6) Cluster $B$ - Groups most susceptible to influence:

Cluster $B$ is most influenced by $E$ and may refer to youth and consumer groups that are easily influenced by social trends, mass media, and influencers. They react quickly to new fads, social movements, or advertising campaigns and often act on their influences.

### (7) Clusters $C$ and $D$ - Intermediate/adaptive groups:

Clusters $C$ and $D$ begin to be influenced over time and remain stable at a certain level. This behavior is characteristic of groups that are open to new information and ideas, but accept them cautiously and incorporate them within existing belief systems and social structures. Examples might include moderate political groups or traditional businesses that gradually adapt to social advances.

### (8) Trends as time-varying:

Short-term fads and campaigns rapidly affect certain populations (e.g., Cluster $B$), but their effects rarely spill over to other populations (Cluster $A$). Some populations (clusters $C$ and $D$) take longer to accept new information and ideas, but once they do, they undergo lasting change. Over time, differences in opinion between groups may narrow or certain groups may become entrenched as opinion leaders.

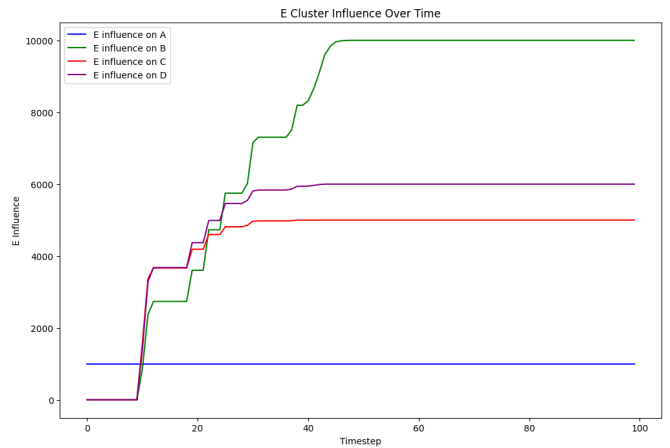

Fig. 5: E Cluster Influence Over Time

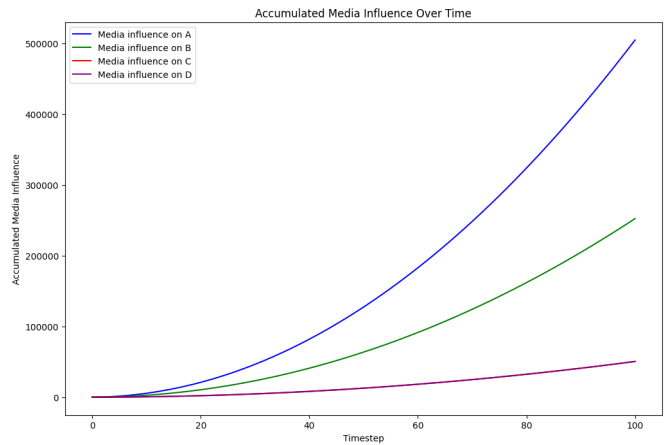

Fig. 6: Accumulated Media Influence Over Time.

**bmedia effect**

| Class | Size | Messages | *bmedia* |
|-------|------|----------|----------|
| A | 100 | 200 | -100 |
| B | 100 | 100 | 100 |
| C | 50 | 100 | 10 |
| D | 60 | 50 | 10 |
| E | 10 | 100 | - |

Table. 2: Class conditions with media influences

Note: Positive media influence (*amedia*) and negative media influence (*bmedia*) are shown to suppress information for each class. Classes without specified media influence are marked with a dash (-).

**(1) Considerations on the behavior of clusters *A-E***

**Clusters with risky behavior**

Cluster *E* is relatively small in number, has a large volume of messages, and has a strong influence on the other clusters. This can be considered a positive risk-taking behavior. If it is small and influential, it may pose a significant risk to other clusters if its information and actions are biased.

**Cluster with one-sided, inductive behavior**

Cluster *E* is considered to be unilaterally guiding because of its influence over all other clusters. The graph shows that its influence on cluster *C* is particularly large.

**Clusters that are acting defensively**

Cluster *A* may have some defensive machinations, as it is not accumulating negative media influences despite the presence of negative media influences.

**(2) Information selection behavior of passive cluster *B***

Cluster *B* is vulnerable to negative media influences, but because the influences build up over time, they need to critically evaluate information and avoid information from negative sources.

Cluster *B* should try to make a balanced selection of information by diversifying information sources and conducting checks to ensure the accuracy of information.

**(3) Consideration of social clusters and patterns of opinion formation**

A pattern of minority opinions influencing the majority can be seen in Cluster *E*. This may be similar to the case in the real world where a minority of experts and activists shape the opinions of the masses.

Clusters *A* and *C* tend to be negatively influenced by the media, but the influences do not stack up. This suggests that these clusters may be media literate and analyze information critically.

**(4) Trends of each cluster based on time trends**

Clusters *B* and *D* show an accumulation of media influence over time, with cluster *B* in particular showing a rapid increase in influence. This indicates the potential for negative information to be reinforced over time and for opinions within a group to become fixed.

Cluster *C* is affected but at a slower pace of increase, suggesting that it either has some mechanism to mitigate the impact or is balancing it by actively incorporating new information.

**(5) Risky Behavior in Speech**

Cluster *E* can be thought of as influential opinion leaders and media entities. For example, news outlets with a specific political leaning or influencers on social networks could correspond here. These entities, though few, may wield significant sway and bring considerable volatility to public discourse.

**Passive Clusters and Information Selection**

Cluster *B* represents the general populace, highly influenced by the daily influx of information yet with little critical filtering, thereby allowing opinions to be molded by persistently negative inputs. For instance, individuals who accept political views or social stereotypes without scrutiny might be categorized under this cluster.

**(6) Social Clusters and Patterns of Opinion Formation**

Patterns in which minority opinions influence the majority might involve a group of experts or thought leaders dictating the societal discourse. An example might be the widespread public endorsement of a scientist's stance on climate change or the swift adoption of an activist movement's agenda.

**(7) Time Trends and Tendencies**

The evolution of clusters *B* and *D* over time exemplifies the entrenchment of opinions due to continuous media influence, akin to the "echo chamber effect", where repeated exposure to congruent information sources solidifies one's thinking in a certain direction.

Cluster *C* could represent communities or groups where education on media literacy and access to varied information is prevalent, as the impacts do not cumulate significantly over time. This suggests an adaptability to new insights and a resistance to one-dimensional viewpoints.

Moreover, given the vast differences dictated by cultural and societal contexts, the examples cited should be taken as generalized case studies.

**kcore effect**

The conditions for the five classes A-E are summarized as follows:

**(1) Clusters with Risky Behavior**

Cluster *B* has a large amount of negative media influence compared to the other clusters, indicating potential social risks.

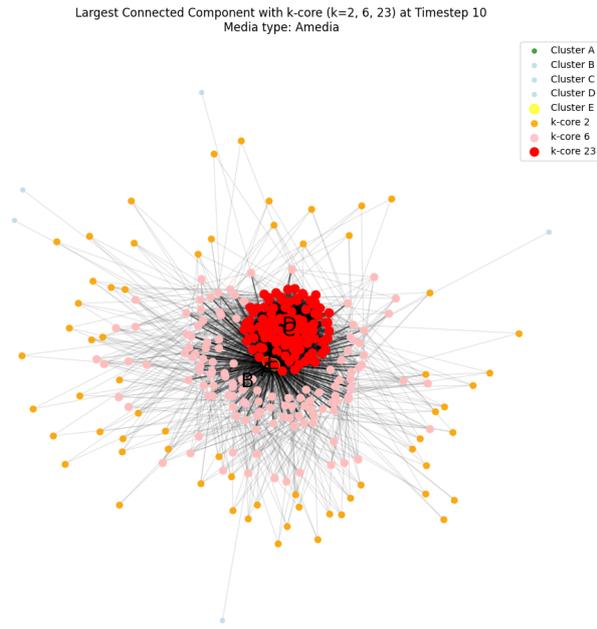

Fig. 7: Largest Connected Component with k-core (k=2, 6, 23) at Timestep $t = 10$

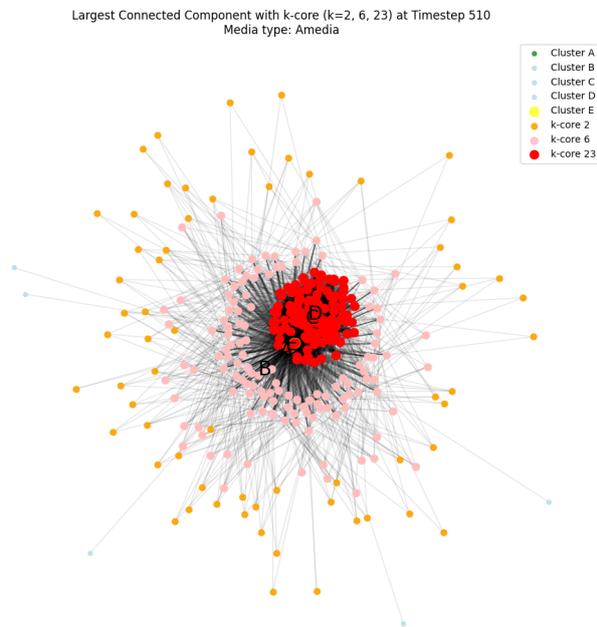

Fig. 8: Largest Connected Component with k-core (k=2, 6, 23) at Timestep $t = 510$

| Class | Size | Messages | *amedia* | *bmedia* |
|---|---|---|---|---|
| A | 100 | 200 | 100 | -10 |
| B | 100 | 100 | 50 | 100 |
| C | 50 | 100 | 10 | 10 |
| D | 60 | 50 | 10 | 10 |
| E | 10 | 100 | - | - |

Table. 3: Class conditions with media influences

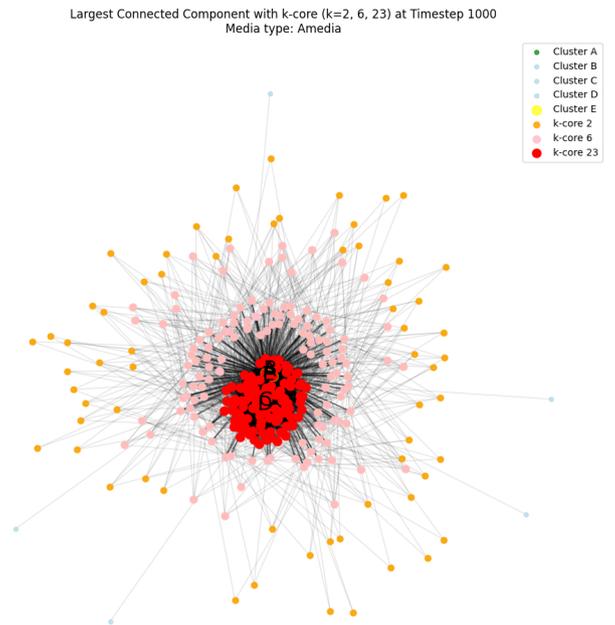

Fig. 9: Largest Connected Component with k-core (k=2, 6, 23) at Timestep $t = 1000$

### (2) Cluster with One-Sided Inductive Behavior

Cluster *A* receives a significant amount of positive media influence and transmits a considerable amount of information, potentially exerting a strong inductive influence on other clusters.

### (3) Clusters with Defensive Behavior

Despite having the largest amount of information, Cluster *E* is the smallest in size, possibly indicating an avoidance of external influences or the formation of a strong inward-looking community.

### (4) Information Selection Behavior of Passive Cluster *B*

Given its susceptibility to negative media influences, Cluster *B* is recommended to diversify information sources and cultivate critical thinking to shield itself from one-sided influences. Actively incorporating positive media influences may also be beneficial in offsetting the negative ones.

**(5)Opinion Leaders**

Cluster *A* likely plays the role of an opinion leader, central to the formation of social opinions.

**(6)Groups Susceptible to Influence**

Cluster *B* is more susceptible to negative influences and thus passive in terms of social opinion formation.

**(7)Niche Communities**

Clusters *C*, *D*, and *E* represent smaller, specialized communities likely engaged in in-depth discussions on specific topics and opinions.

**Trend of Each Cluster Based on Time Step**

Over time, Cluster *A* maintains its central role in opinion formation. By time step 1000, the centrality of Cluster *B* declines, suggesting a reduction in influence due to negative media impacts. The consistent positions of Clusters *C*, *D*, and *E* from time step 100 to 1000 indicate a steady adherence to certain niche opinions and values.

**(8)Cases of Real Social Clusters**

**Cluster *A***: This includes mainstream media and political leaders as "opinion leaders" and influencers on major social networking sites. They exert positive influence on a wide range of followers and set social trends.

**Cluster *B***: The "susceptible group" includes consumers who rely on specific sources of information and those who are strongly influenced by a particular ideology. This group is more likely to have its opinions shaped by fake news and propaganda.

**Clusters *C*, *D*, *E***: As for "niche communities," these are small online forums or communities that share a hobby, profession, field of study, or specific ideology. They are deeply immersed in a particular topic or interest and have close social ties.

**(9)Hypothesis of Trends as Time-Varying**

**Short-term trend**: When new information or trends emerge, Cluster *A* is quick to pick up on and disseminate it. Cluster *B* is more receptive to this information, but may lack the ability to verify it and risk spreading misinformation.

**Medium-term trend**: While opinions are formed under the influence of Cluster *A*, Clusters *C*, *D*, and *E* continue to hold their own opinions, thus preserving diversity in society as a whole. However, the negative influence of Cluster *B* may gradually erode social trust.

**Long-term trend**: When Cluster *A* has persistent influence, social norms and mainstream opinion are stable. However, if Cluster *B* is under constant negative influence, social fragmentation may deepen. Clusters *C*, *D*, and *E* may contribute to social progress by developing new subcultures and specialized areas of knowledge.

**(10)Practical Examples**

**Social movements**: A social movement on social media (e.g., the #MeToo movement) could spread rapidly with the support of influencers and media in Cluster *A*. However, Cluster *B* could be negatively impacted by manipulation of information by those who oppose the movement.

**Political Opinion**: During elections, the media and politicians in Cluster *A* have the power to shape public opinion. Cluster *B* is strongly influenced by them, while clusters *C*, *D*, and *E* may hold more professional or individualized opinions.

**Technological advances**: Information about new technologies (e.g., electric cars, renewable energy) is first spread by Cluster *A* and followed by Cluster *B*, while Clusters *C*, *D*, and *E* may explore the specialized aspects of the technology and deepen their practical knowledge.

# Aknowlegement

This research is supported by Grant-in-Aid for Scientific Research Project FY 2019-2021, Research Project/Area No. 19K04881, "Construction of a new theory of opinion dynamics that can describe the real picture of society by introducing trust and distrust". It is with great regret that we regret to inform you that the leader of this research project, Prof. Akira Ishii, passed away suddenly in the last term of the project. Prof. Ishii was about to retire from Tottori University, where he was affiliated with at the time. However, he had just presented a new basis in international social physics, complex systems science, and opinion dynamics, and his activities after his retirement were highly anticipated. It is with great regret that we inform you that we have to leave the laboratory. We would like to express our sincere gratitude to all the professors who gave me tremendous support and advice when We encountered major difficulties in the management of the laboratory at that time.

First, Prof. Isamu Okada of Soka University provided valuable comments and suggestions on the formulation of the three-party opinion model in the model of Dr. Nozomi Okano's (FY2022) doctoral dissertation. Prof.Okada also gave us specific suggestions and instructions on the mean-field approximation formula for the three-party opinion model, Prof.Okada's views on the model formula for the social connection rate in consensus building, and his analytical method. We would also like to express our sincere gratitude for your valuable comments on the simulation of time convergence and divergence in the initial conditions of the above model equation, as well as for your many words of encouragement and emotional support to our laboratory.

We would also like to thank Prof.Masaru Furukawa of Tottori University, who coordinated the late Prof.Akira Ishii's laboratory until FY2022, and gave us many valuable comments as an expert in magnetized plasma and positron research.

In particular, we would like to thank Prof.Hidehiro Matsumoto of Media Science Institute, Digital Hollywood University. Prof.Hidehiro Matsumoto is Co-author of this paper, for managing the laboratory and guiding us in the absence of the main researcher, and for his guidance on the elements of the final research that were excessive or insufficient with Prof.Masaru Furukawa.

And in particular, Prof.Masaru Furukawa of Tottori University, who is an expert in theoretical and simulation research on physics and mathematics of continuum with a focus on magnetized plasma, gave us valuable opinions from a new perspective.

His research topics include irregular and perturbed magnetic fields, MHD wave motion and stability in non-uniform plasmas including shear flow, the boundary layer problem in magnetized plasmas, and pseudo-annealing of MHD equilibria with magnetic islands.

We received many comments on our research from new perspectives and suggestions for future research. We believe that Prof.Furukawa's guidance provided us with future challenges and perspectives for this research, which stopped halfway through. We would like to express sincere gratitude to him.

We would like to express my sincere gratitude to M Data Corporation, Prof.Koki Uchiyama of Hotlink Corporation, Prof.Narihiko Yoshida, President of Hit Contents Research Institute, Professor of Digital Hollywood University Graduate School, Hidehiko Oguchi of Perspective Media, Inc. for his valuable views from a political science perspective. And Kosuke Kurokawa of M Data Corporation for his support and comments on our research environment over a long period of time. We would like to express our gratitude to Hidehiko Oguchi of Perspective Media, Inc. for his valuable views from the perspective of political science, as well as for his hints and suggestions on how to build opinion dynamics.

We are also grateful to Prof.Masaru Nishikawa of Tsuda University for his expert opinion on the definition of conditions in international electoral simulations.

We would also like to thank all the Professors of the Faculty of Engineering, Tottori University. And Prof.Takayuki Mizuno of the National Institute of Informatics, Prof.Fujio Toriumi of the University of Tokyo, Prof.Kazutoshi Sasahara of the Tokyo Institute of Technology, Prof.Makoto Mizuno of Meiji University, Prof.Kaoru Endo of Gakushuin University, and Prof.Yuki Yasuda of Kansai University for taking over and supporting the Society for Computational Social Sciences, which the late Prof.Akira Ishii organized, and for their many concerns for the laboratory's operation. We would also like to thank Prof.Takuju Zen of Kochi University of Technology and Prof.Serge Galam of the Institut d'Etudes Politiques de Paris for inviting me to write this paper and the professors provided many suggestions regarding this long-term our research projects.

We also hope to contribute to their further activities and the development of this field. In addition, we would like to express our sincere gratitude to Prof.Sasaki Research Teams for his heartfelt understanding, support, and advice on the content of our research, and for continuing our discussions at a time when the very survival of the research project itself is in jeopardy due to the sudden death of the project leader.

We would also like to express our sincere gratitude to the bereaved Family of Prof.Akira Ishii, who passed away unexpectedly, for their support and comments leading up to the writing of this report. We would like to close this paper with my best wishes for the repose of the soul of Prof.Akira Ishii, the contribution of his research results to society, the development of ongoing basic research and the connection of research results, and the understanding of this research project.